# Towards a Climate OSSE Framework for Satellite Mission Design

Ann M. Fridlind,[a] Gregory S. Elsaesser,[a,b] Marcus van Lier-Walqui,[a,c] Grégory V. Cesana,[a,c] Elizabeth Weatherhead,[d] George Tselioudis,[a] Gavin Schmidt,[a] Donifan Barahona,[e] Brian Cairns,[a] William D. Collins,[f] David Considine,[g] Lidia Cucurull,[h] Larry DiGirolamo,[i] Amber Emory,[g] Otto Hasekamp,[j] Shan He,[k] Ryan Kramer,[l] Matthew Lebsock,[m] Tsengdar Lee,[g] Stephen Leroy,[n] Wuyin Lin,[o] Steven Lugauer,[p] Daniel Miller,[e,q] Johannes Mülmenstädt,[r] Lazaros Oreopoulos,[e] Derek J. Posselt,[m] and Mark D. Zelinka[s]

[a] *NASA Goddard Institute for Space Studies, New York, NY*

[b] *Department of Applied Physics and Applied Mathematics, Columbia University, New York, NY*

[c] *Center for Climate Systems Research, Columbia University, New York, NY*

[d] *University of Colorado, Boulder, CO*

[e] *NASA Goddard Space Flight Center, Greenbelt, MD*

[f] *Lawrence Berkeley National Laboratory, Berkeley, CA*

[g] *NASA Headquarters, Washington, DC*

[h] *NOAA Environmental Modeling Center, College Park, MD*

[i] *University of Illinois at Urbana-Champaign, Urbana, IL*

[j] *Netherlands Institute for Space Research, Leiden, Netherlands*

[k] *Stony Brook University, Stony Brook, NY*

[l] *Geophysical Fluid Dynamics Laboratory, Princeton, NJ*

[m] *Jet Propulsion Laboratory, California Institute of Technology, Pasadena, CA*

[n] *Atmospheric and Environmental Research, Lexington, MA*

[o] *Brookhaven National Laboratory, Brookhaven, NY*

[p] *University of Kentucky, Lexington, KY*

[q] *University of Maryland, Baltimore County, MD*

[r] *Pacific Northwest National Laboratory, Richland, WA*

[s] *Lawrence Livermore National Laboratory, Livermore, CA*

*Corresponding author*: Ann M. Fridlind, ann.fridlind@nasa.gov

## ABSTRACT

The rich history of observing system simulation experiments (OSSEs) does not yet include a well-established framework for using climate models. The need for a climate OSSE is triggered by the need to quantify the value of a particular measurement for reducing the uncertainty in climate predictions, which differ from numerical weather predictions in that they depend on future atmospheric composition rather than the current state of the weather. However, both weather and climate modeling communities share a need for motivating major observing system investments. Here we outline a new framework for climate OSSEs that leverages the use of machine-learning to calibrate climate model physics against existing satellite data. We demonstrate its application using NASA's GISS-E3 model to objectively quantify the value of potential future improvements in spaceborne measurements of Earth's planetary boundary layer. A mature climate OSSE framework should be able to quantitatively compare the ability of proposed observing system architectures to answer a climate-related question, thus offering added value throughout the mission design process, which is subject to increasingly rapid advances in instrument and satellite technology. Technical considerations include selection of observational benchmarks and climate projection metrics, approaches to pinpoint the sources of model physics uncertainty that dominate uncertainty in projections, and the use of instrument simulators. Community and policy-making considerations include the potential to interface with an established culture of model intercomparison projects and a growing need to economically assess the value-driven efficiency of social spending on Earth observations.

## SIGNIFICANCE STATEMENT

When planning a new satellite mission, it is important to first make sure that the new measurements will meet the science and end user goals of the broader community. While there are now well-established ways to quantify observation benefits for weather prediction, there is no such similar established framework for determining satellite measurement benefits for climate prediction. This article describes a new way to determine in advance whether new observations can reduce uncertainty in climate model projections.

## CAPSULE (BAMS only)

A new type of observing system simulation experiment quantifies the benefit of new satellite missions for climate modeling and projection.

# 1. Introduction

A well-established framework exists for observing system simulation experiments (OSSEs) that quantify the value of future observations for improving numerical weather prediction (NWP) skill, which in the US is now centralized at NOAA's Quantitative Observing System Assessment Program (QOSAP). By contrast, such a capability does not currently exist for climate OSSEs that rely on Earth system models (ESMs), either in the US or internationally.

An OSSE can be broadly defined as any experiment that quantifies the value of an observing system (Weatherhead et al. 2018). Past efforts to establish a comprehensive climate OSSE capability at NASA offered the following distinctions (Leroy et al. 2016): '*A climate OSSE is a statistical computation based upon arbitrarily complex models of the climate that objectively evaluates mission requirements, instrument requirements, and measurement requirements for proposed observing systems and quantifies the value an observing system adds to testing well-defined hypotheses or to improving climate prediction*.' In a review of OSSE use across US agencies tasked with environmental observations, Zeng et al. (2020) describe NASA's use of *sampling* OSSEs to evaluate temporal and spatial sampling and *retrieval* OSSEs to quantify information on a geophysical quantity, and recommend that OSSE development for ESMs be accelerated and extended to assess societal impacts.

A leading application of a mature climate OSSE capability would be to systematically quantify the capabilities of future Earth observing satellites to meet mission objectives related to climate prediction, such as those defined in the most recent US decadal survey for Earth science and applications from space (ESAS; NASEM 2018). Increasing levels of quantitative traceability have already been introduced into NASA's protocols for mission formulation. A notable example is mandatory use of a standardized science and applications traceability matrix (SATM), which demonstrates traceability from mission architecture to science questions. The ESAS decadal survey's development process for missions in the committed designated observable (DO) class further implemented a quantitative framework in the downselection process during mission formulation. Specifically, the Aerosol Clouds Convection Precipitation (ACCP) DO mission implemented a value framework through which to quantitatively compare the capability of candidate observing system architectures to address the science question enumerated in an SATM (Ivanco and Jones 2020). The science value is calculated as the product of two components: the quality (how accurately a geophysical variable is measured) and the utility (how useful that geophysical variable is in addressing the science questions). The introduction of this quantitative framework in the architecture down-selection process is an advance, but notable gaps remain in developing a fully quantitative traceability from observing system architectures to addressing the climate science questions articulated in the ESAS decadal survey. For example, a specific objective of the climate panel (NASEM 2018; objective C2a) is to '*Reduce uncertainty in low and high cloud feedback by a factor of 2*.' The value framework, as formulated, provides no direct means of assessing if a mission architecture could achieve this objective. For this purpose a new framework leveraging new tools is required.

For NWP models, the value of proposed observing systems can be compared using established forecast skill metrics, often using a high-resolution reanalysis "nature run" as a source of synthetic observations. For instance, to evaluate future atmospheric sounding observing system capabilities (e.g., orbit configuration, technology choice, horizontal resolution, and revisit rate), Cucurull et al. (2024) conducted an NWP OSSE using QOSAP's consolidated observing system simulator (COSS) package with a 9-km ECMWF "nature run" to generate synthetic data sets for assimilation by the NOAA Global Forecast System (GFS). By contrast, the diverse goals of climate observing systems lead to a diversity of evaluation approaches. For instance, Chepfer et al. (2018) used two climate models to estimate how the length and continuity of multidecadal spaceborne lidar records would impact their capability to constrain regional cloud feedbacks. However, an objective of many shorter Earth observing satellite missions is to directly reduce uncertainty in ESM physical processes rather than emergent quantities such as cloud feedbacks, which motivates this work.

Here we introduce a new climate OSSE approach for quantifying how improved satellite retrievals can better constrain ESM physics parameters, which then can be traced through to a change in climate projection envelopes. As a proof-of-concept exercise, we quantify the degree to which a potential NASA Planetary Boundary Layer (PBL) mission (Teixiera et al. 2025) could better constrain NASA's GISS-E3 model and its climate projections. We note that the technologies under consideration for the PBL mission (cf. Teixeira et al. 2025) have already been examined in a retrieval OSSE framework (Kurowski et al. 2023) and overlap significantly with those compared in the Curcurull et al. (2024) NWP OSSE. After outlining a schematic workflow (Section 2), we discuss technical considerations (Section 3) and community and policy considerations (Section 4).

## 2. A proof-of-concept

NWP and data assimilation (DA) primarily focus on using Earth observations to reduce uncertainty in the state of the atmosphere for initializing forecasts. In contrast to initial-value predictability (Palmer and Hagedorn 2006), modeling of the Earth's climate on timescales of years, decades, and centuries presents a different challenge: boundary-value or forced predictability (predictability of the second kind; Lorenz 1975). For future climate simulations, in addition to uncertainty arising from emission scenarios (Lehner et al. 2020), a primary concern is uncertainty arising from the underlying numerical representation of physical processes. This includes methods of discretization and time-stepping, as well as the parameterization of many interacting and unresolved sub-grid processes, which for simplicity we collectively refer to hereafter as "model physics". The information content of observations for climate modeling is assessed relative to how it constrains the envelope of projections that tie to uncertain model physics rather than how it informs an uncertain state. Below, we outline an ESM-based OSSE workflow that enables quantifying the reduction in climate prediction uncertainty due to proposed observations. Via use of NASA's GISS-E3 ESM in the workflow, we then demonstrate a specific example of how improved satellite retrievals of near-surface atmospheric state could further constrain uncertain ESM physics parameters and, in turn, impact projection envelopes.

ESM model physics uncertainty can be conceptually separated into two sources: parametric uncertainties, and structural uncertainties. The former is associated with parameters in model physics parameterizations schemes, for example a coefficient that controls snow crystal fall speed, and the latter is related to structural choices that cannot be easily adjusted by changing parameter values, for example the choice to represent cloud droplet size distributions with a gamma distribution. Here we begin with a focus on parametric uncertainty, and later we discuss how this proof-of-concept can be extended to consideration of structural uncertainties (Section 3). Observational constraint of parameters faces unique challenges relative to constraint of initial model state: while the number of uncertain model physics parameters ($<10^3$) is typically much less than that of state variables ($>10^6$), the relationship between observable quantities and parameters can be highly nonlinear, ruling out methods such as ensemble Kalman filters and variational techniques. Instead, methods such as Markov chain Monte Carlo (MCMC; Posselt 2016; van Lier-Walqui et al. 2014) are typically used, requiring many thousands or millions of parameter samples, each with corresponding computationally intensive model forecasts. As in other fields of physical science facing similar challenges (e.g., Kasim et al. 2021), this cost has motivated machine learning (ML) methods for emulating the sensitivity of ESMs to parameter perturbations (e.g., Watson-Parris et al. 2021; Eidhammer et al. 2024), where such emulators can then stand in for the full ESM in MCMC or other computationally demanding inference approaches (e.g., Elsaesser et al. 2025). With these methods in hand, uncertainty can be quantified in model parameters, strictly informed by observational uncertainty, via the formalism of Bayesian inference. Such methods are a prerequisite for the ESM-based climate OSSE approach introduced here.

First, in order to accurately quantify improved constraints attributed to new observations, it is crucial that the current observational record is maximally utilized in the development of a baseline set of constraints (i.e., reference constraints). In the absence of being able to observe all physical processes important to ESM projection fidelity, using as many diverse observations as possible that are *currently* available (beyond those most commonly used in many climate model calibration efforts; e.g., Schmidt et al. 2017) provides a pathway to development of baseline constraints. The set of observations could include information on convective surface rainfall rate distributions, stratocumulus and shallow cumulus cloud fractions, cloud top phase statistics, etc., alongside the more traditional set of observations (mean precipitation rate, water vapor, circulation diagnostics, low and total cloud fraction, radiation field, tropospheric water vapor, and temperature). Utilizing a diverse set of existing satellite observations of the atmospheric state (which then comprises the set of baseline constraints) is important since the suite of cloud feedbacks might be more strongly related to the base state if more comprehensively defined, and more generally, we may become closer to simulating Earth-like climatologies for the right process-level reasons (for example, mitigating the common “too few, too bright” bias in ESM cloud properties; Nam et al. 2012).

An expanded set of baseline constraints enables derivation of a more complete set of plausible model physics parameter combinations (i.e., a “calibrated” parameter set), where each parameter combination yields model climatologies that approximately match all baseline constraints. Finding likely physics parameter combinations requires an improved assessment of the uncertainty that exists in observational data products (Elsaesser et al. 2025), and thus, maximally utilizing all available observations only works well if the observational uncertainties are well characterized. Observational uncertainty widens the state space of physics parameters whose combinations yield ESM configurations in agreement with an expanded set of (uncertain) observational constraints. Quantifying overall observational uncertainty, and using it to determine the plausible model physics parameter combinations is the second key component enabling our proof-of-concept, given that it relates to setting measurement requirements as part of an OSSE. Each ESM physics parameter combination in turn maps to a scenario-dependent prediction of future climate features, such as equilibrium climate sensitivity (surface air warming resulting from a benchmark doubling of carbon dioxide concentration), extreme event occurrences, and the evolution of a broad spectrum of climatic impact drivers (CIDs; Ruane et al. 2022) affecting human and natural systems. The aggregate of all calibrated ESM physics parameter combinations finally yields a baseline prediction envelope against which we can now objectively evaluate whether any potential new observations will result in refinement (narrowing, shifting) of the predictions.

The components described above are part of a methodology for estimating all plausible ESM physics parameters leading to ESM simulations that are in good agreement with existing satellite observations of the atmospheric state, resulting in what we call a calibrated physics ensemble (CPE; Elsaesser et al. 2025). The posterior set of “good” physics parameters, derived from an MCMC sampler that probabilistically samples parameters in accordance with the misfit between the emulator outputs and all observations (within their uncertainties), are then tested in climate model simulations to verify that they indeed yield viable ESM configurations; if they do not, they are discarded from CPE membership. In this way, the CPE becomes a collection of true climate model configurations, emerging from the collection of "good" physics parameter combinations produced from the emulator-MCMC methodology. Model-observation misfits caused by systematic (or structural) errors in ESM parameterizations are accounted for in this process, which is important for simultaneous use of numerous observations; otherwise, a large misfit overwhelms the ability to optimize all physics parameters (see details in Elsaesser et al. 2025). The collection of plausible ESM variants comprising the CPE map to an envelope of projections, as illustrated in the top row of the Fig. 1 schematic.

Importantly, the surrogate model is not limited to emulating only outputs for which an observational data set currently exists. Using a specific CPE created from the NASA Goddard Institute for Space Studies ESM (i.e., the 'GISS-E3 CPE'; Elsaesser et al. 2025), we can now provide one demonstration of what new observations and uncertainty thresholds might do to constrain physics parameters and projection envelopes.  For example, consider the current problem of remotely sensing the cloudy planetary boundary layer (PBL). At its best, the PBL temperature and water vapor from the existing observational record is largely representative of cloud-free conditions. At its worst, there might be large biases due to the difficulty of retrieving the thermodynamic state of the PBL with current satellite infrared and microwave sounder channel selections and weighting functions. In the GISS-E3 CPE, water vapor at 925 hPa (qv925) was used as one observational constraint among 36 data sets (full list in Elsaesser et al. 2025).  The associated estimate of uncertainty in the observation of qv925 is also relatively large, insofar as the constraint it provides is relatively weak. We can infer this by systematically increasing the specified uncertainty in qv925, re-estimating the "good" posterior physics parameter combinations, and finding that they are negligibly different from the baseline set (Fig. 2a). By extension, there is no expected change in a full CPE projection envelope.

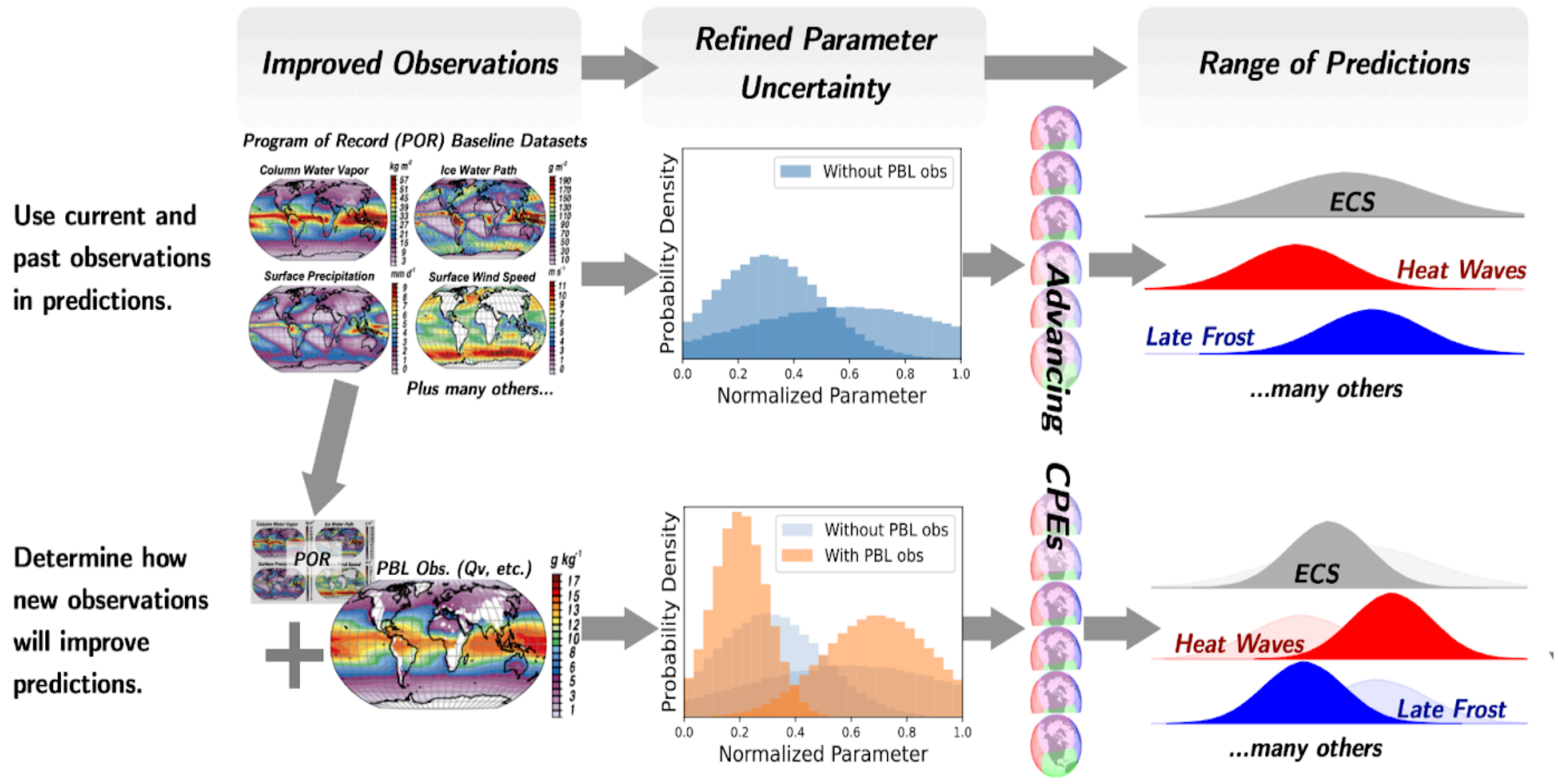

Fig. 1. Schematic illustrating the use of program of record (POR) datasets and new PBL observations to constrain and narrow the distribution of plausible physics parameter settings in an ESM, which results in a progressively-constrained set of ESM calibrated physics ensemble (CPE) configurations that translate to refined ranges of predictions (e.g., equilibrium climate sensitivity (ECS), heat wave, and late frost projections).

However, if we systematically decrease uncertainty, which is analogous to development and incorporation of an improved qv925 product, we find that some physics parameter histograms narrow and/or shift (Fig. 2b), while others do not. This is the desired response from a methodology that can serve as a climate OSSE framework. In other words, we generally expect enhanced PBL observations to constrain parameters of processes closely linked to the PBL – for example, a strongly affected parameter "bsort_enteff2" is the entrainment rate for the more-entraining plume in the GISS-E3 convection scheme (a process directly tied to PBL relative humidity), whereas a less affected parameter "dcs" relates to the autoconversion of cloud ice to snow in stratiform clouds, largely impacting simulated ice water mass in clouds above the melting level (a property we expect to be less affected by PBL characteristics).

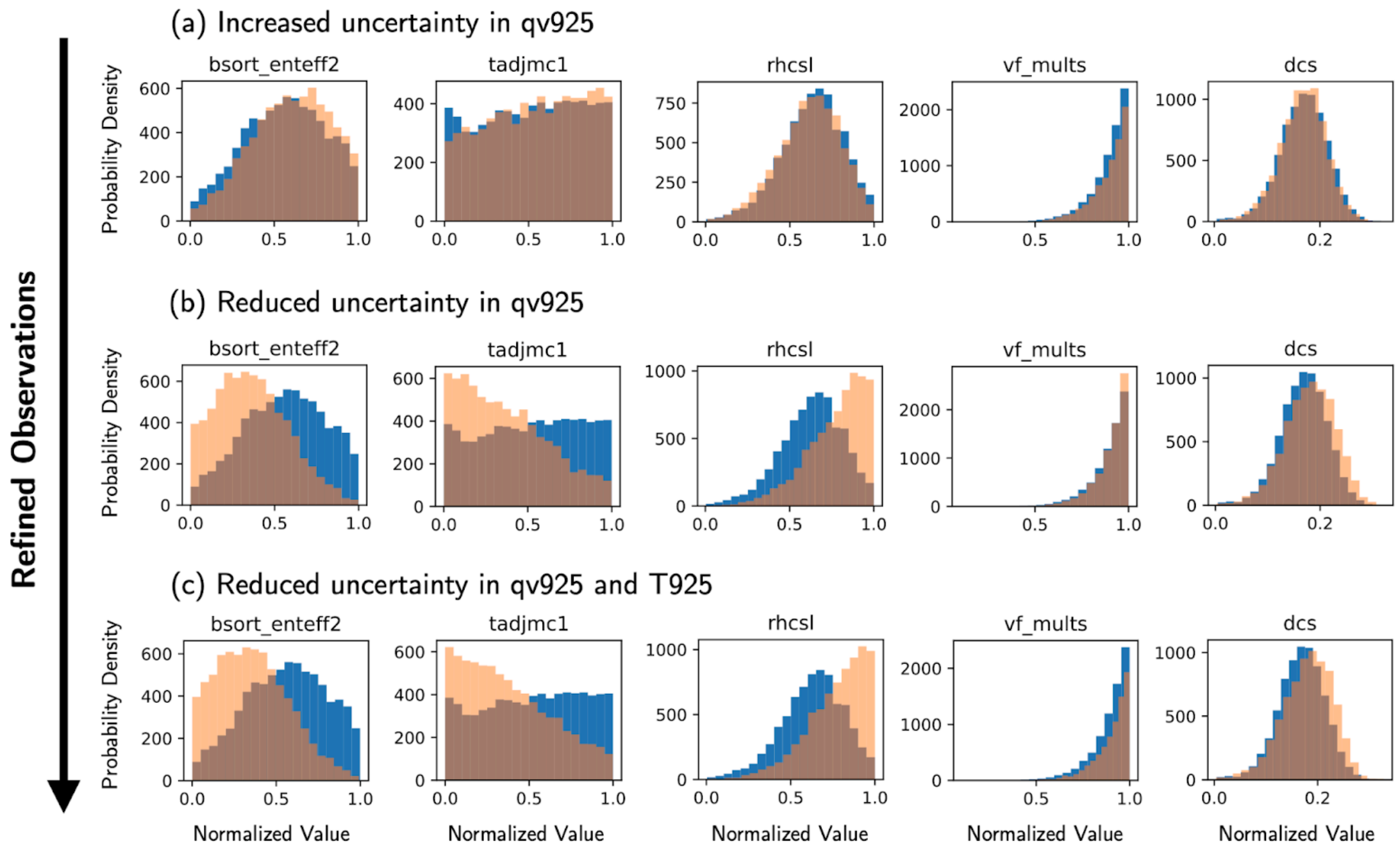

.

Fig. 2. Posterior probability densities as a function of the normalized range of values of five physics parameters in the GISS-E3 baseline set (blue) versus a proof-of-concept climate OSSE set (orange) with a factor of five (a) increased uncertainty in qv925, (b) reduced uncertainty in qv925, and (c) reduced uncertainty in both qv925 and T925. Parameters shown are the cumulus entrainment rate (bsort_enteff2), cumulus relaxation timescale (tadjmc1), critical relative humidity for stratiform cloud formation (rhcsl), stratiform snow sedimentation rate (vf_mults), and a critical diameter for autoconversion of cloud ice to snow in stratiform clouds (dcs).

We can now trace the impact of shifting and narrowing parameter histograms through to impact on projections. To demonstrate this, we built a GISS-E3 net cloud feedback emulator based on the relationship between variability in physics parameters and net cloud feedback for a subsample of 54 GISS-E3 CPE members. The net cloud feedback is computed using cloud radiative kernels from Zelinka et al. (2012) with 5-year long AMIP and AMIP-p4K simulations. We additionally compute ECS estimates using predicted cloud feedbacks and a linear relationship between net cloud feedback and ECS (e.g., Zelinka et al. 2017, 2020), derived from 15 GISS-E3 CPE members of the larger 54-member ensemble.  These 15 were those that exhibited similar and realistic TOA radiative imbalances required to run a slab ocean model with 4x$CO_2$ perturbations and derive climate sensitivity estimates (Gregory et al. 2004). Note that the non-cloud feedbacks are fairly similar for these 15 configurations and that we assume that the relationship between net cloud feedback and ECS holds across all members. Using this emulator to derive net cloud feedback from the new qv925-constrained GISS-E3 CPE ensemble shows a substantial narrowing of the spread in net cloud feedback compared to the control CPE (Fig. 3, left), which, ultimately, translates into a reduction of the climate sensitivity spread (Fig. 3, right).

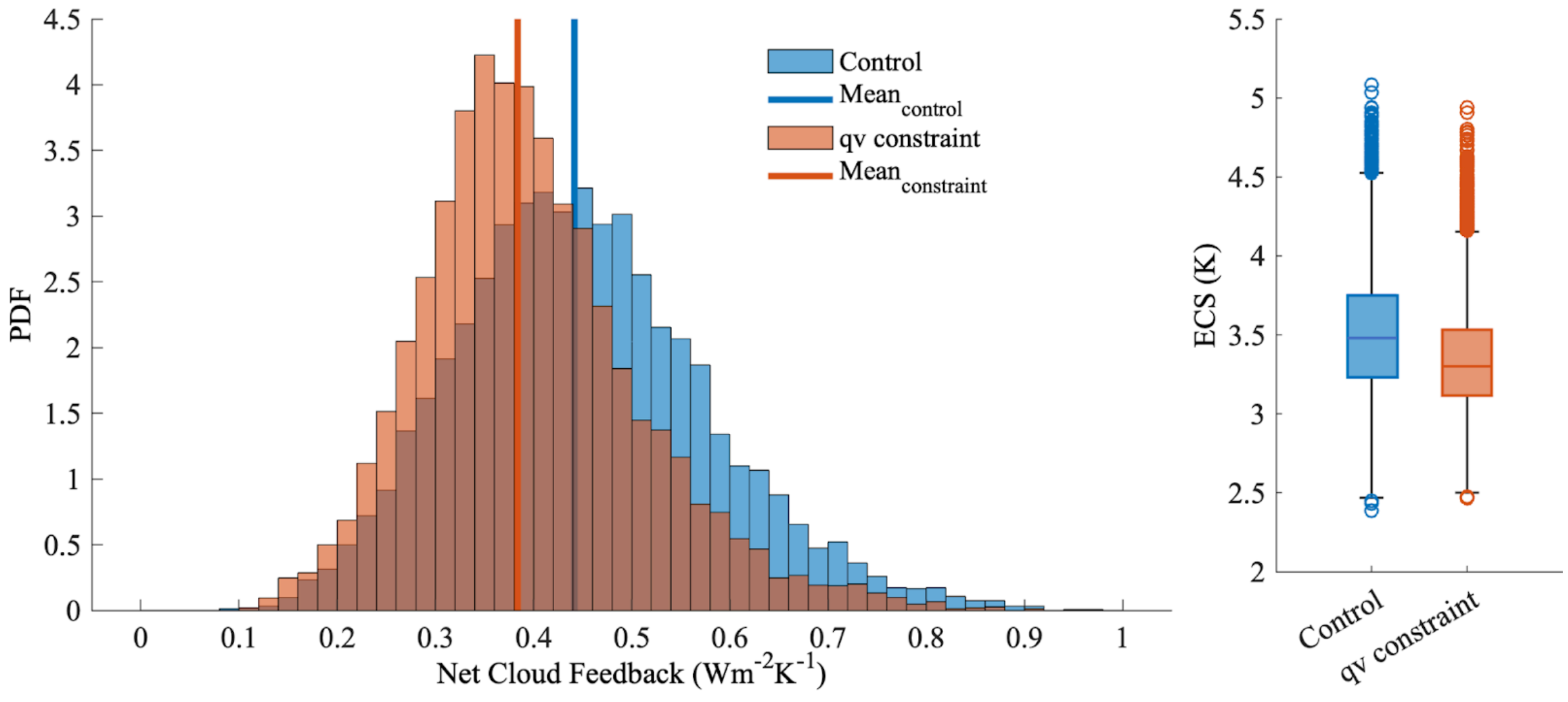

Fig. 3: Reducing uncertainty in qv925 reduces the spread (and mean value) of cloud feedbacks as well as climate sensitivity estimates (right) in the refined GISS-E3 CPE (red, n = 7500; left) compared to the baseline GISS-E3 CPE (blue control, n = 7500). The box chart (right) shows the median, first and third quartiles, outliers (circles, computed using the interquartile range), and minimum and maximum values that are not outliers. See manuscript text for explanation of net cloud feedback and ECS calculation.

This specific qv925 constraint – refined projection demonstration can be generalized to the hypothetical constraint concept reflected in the bottom row of the Fig. 1 schematic. Extending the proof-of-concept demonstration further, a climate OSSE can reveal the relative impact of different observations: in row (c) of Fig. 2, we repeat the experiment of row (b) but also reduce uncertainty in temperature at 925 hPa (T925). Inclusion of reduced uncertainty in temperature is shown to have very little (nearly negligible) impact relative to reduction in uncertainty in qv925, indicating that accuracy of boundary layer humidity observations may be more critical to optimize compared with temperature. It is worth noting that these results are consistent with Suselj et al. (2020) who also used a Bayesian framework to identify the importance of water vapor measurements relative to temperature in constraining single-column model physics and further showed the trade space between measurement vertical resolution and measurement accuracy.

This proof-of-concept example of a climate OSSE approach is a first, relatively simple, step toward a framework that assesses the potential value of planned observational platforms for constraining ESM physics and physics-dependent projections at climatic timescales. We next address various technical aspects relevant to establishing a more mature framework.

## 3. Technical considerations

As noted above, the OSSE framework discussed thus far is limited to parametric uncertainty in model physics. Structural uncertainty, on the other hand, arises from insufficiencies in the interconnected formulation of parameterizations and other factors such as model resolution. Resolving parametric uncertainties within a given structural implementation may further be a prerequisite for effective identification and reduction of structural uncertainties (e.g., Smalley et al. 2023). One way to extend the framework to encompass structural uncertainty would be to span multiple climate models with diverse structural formulations. Information gain that is found to be consistent across diverse ESMs can then be considered more robust.

The OSSE framework demonstrated thus far also remains essentially a black box. Results in Section 2 illustrate that improved observations could better constrain some GISS-E3 physics parameters and climate projections, but what processes changed? Because models producing plausible atmospheric states may result from counteracting biases among different processes (e.g., Kim and Jin 2011), networks that represent climate component interactions might serve as a more effective foundation for quantifying model-observation correspondence. In particular, Bayesian networks offer a viable solution to provide robust causal interpretations (Peters et al. 2017; Runge et al. 2019) and associated uncertainty quantification using sampling methods (e.g., He et al. 2023; O'Kane et al. 2024). Bayesian network metrics based on linear Gaussian models, despite some limitations, have been demonstrated to effectively constrain the uncertainty in climate model projections when used as projection weights (Nowack et al. 2020; Ricard et al. 2024). For instance, Ricard et al. (2024) uses Bayesian network metrics derived from present-day sea surface temperature to evaluate model capabilities to represent real-world physical processes. Accordingly, weights are assigned to the spreads of ECS and transient climate response of ESMs, reducing the range of these spreads. Such an approach links the climate projections to the physics parameter candidates selected based on the network metrics, suggesting that Bayesian networks could offer an effective pathway to including process-oriented causal interpretation within the OSSE framework.

The GISS-E3 CPE and our proof-of-concept OSSE framework operate entirely on the present-day atmospheric state, but the metric that should ultimately be optimized is climate projection uncertainty, illustrated here in terms of simulated cloud feedbacks and ECS. The selection of a subset of metrics, used both to objectively select CPE candidates and to identify prediction uncertainty, can become a cumbersome task if the number of potential metrics grows to accommodate all relevant observables and spatiotemporal scales. An optimum subset of metrics can be better defined if the purpose of the climate OSSE exercise is specified in enough detail to allow for the selection of the quantities that most closely fit the purpose. When the OSSE exercise is used in a satellite mission design process, then the mission requirements can be used as the baseline to define the OSSE's purpose. Ideally the mission objectives identify targets for reducing a specific projection uncertainty, such the halving of low and high cloud feedbacks targeted by the ESAS decadal survey.

For any selected metrics, our OSSE framework relies on the existing observational record to define baseline constraints, and a key aspect that serves as a basis for our proof-of-concept exercise is the data uncertainty, which is often poorly quantified. Systematic errors arise when assumptions made by a retrieval algorithm do not apply in nature. For example, in retrieving cloud optical depth and droplet effective radius from measurements of scattered sunlight, the sub-pixel cloud field is assumed to be spatially homogeneous (e.g., Platnick et al. 2003). But real clouds deviate from this assumption (Di Girolamo et al. 2010), leading to systematic errors that co-vary with spatial heterogeneity and sun-view geometry (e.g., Liang and Di Girolamo 2013; Fu et al. 2019) and to artificial space-time variability in data records (Di Girolamo et al. 2010). Another contribution to systematic error is lack of representativeness when instantaneous measurements are aggregated in time and space to compile climatologies (e.g., Posselt et al. 2012), which are inputs to the CPE and OSSE. For instance, instantaneous measurements are often preferential to particular conditions, e.g., towards large clear sky areas for PBL profiles from IR sounders. Owing to a common lack of comprehensive uncertainty quantification overall, the GISS-E3 CPE resorted to using more than one independent data source for all geophysical quantities (cf. Elsaesser et al. 2025). However, this could underestimate uncertainty if datasets share similar systematic biases, or overestimate uncertainty if their sampling characteristics vary widely. We note that the proof-of-concept procedure in Section 2 could also serve to quantify the degree to which improved uncertainty quantification affects model physics calibration.

Finally, a climate OSSE can be viewed as a further enhancement to retrieval OSSEs that are already in use for satellite mission evaluation (Miller et al. 2018; Castellanos et al. 2019; Liu and Mace 2022; Posselt et al. 2022). Retrieval OSSEs couple synthetic observables of a prospective satellite instrument (e.g., reflectances) with retrieval algorithms (e.g., bispectral retrievals of cloud microphysical quantities) to quantify the impact of instrument design decisions (e.g., spectral bandpass selection) on retrieved geophysical variables. Spatial and temporal sampling characteristics are further considerations. For example, future hyperspectral infrared sounding instruments could offer global hourly clear-sky profiles at coarse vertical resolution, whereas radio occultation observations could offer high vertical resolution at coarse horizontal resolution and sparse temporal coverage. OSSEs that quantify how diverse measurement characteristics can inform the estimates of geophysical quantities (Kurowski et al. 2023) can be connected to the climate OSSE framework by identifying an input to the CPE. If the retrieval OSSE generates a geophysical variable contained in the CPE framework (e.g., qv925 in our proof-of-concept; at lower left in Fig. 1), then passing it through the Bayesian CPE framework provides a means of linking instrument, retrieval and sampling characteristics to improved climate metric constraints (at lower right in Fig. 1). Alternatively, an instrument simulator approach may be used to mimic what a spaceborne instrument orbiting above an ESM atmosphere would observe (Bodas-Salcedo et al. 2011; Cesana and Chepfer 2013; Pincus et al. 2012; Webb et al. 2001), enabling the observations to be used directly as a CPE input.

## 4. Community and policy-making considerations

What triggers a climate OSSE need? The climate OSSE framework demonstrated here is specifically responsive to a call for introducing the use of climate models into PBL mission incubation (Teixeira et al. 2025), where there is a need for comparing measurement value across widely diverse architectures (e.g., technological approaches enumerated by Teixeira et al. 2025).

Teixeira et al. (2021) note that prediction of societally impactful atmospheric circulation features such as mid-latitude blocking events is sensitive to PBL physics in climate models (e.g., Lindvall et al. 2017), and it is fundamentally unknown to what degree improved PBL measurements over particular regions such as the Southern Ocean could reduce uncertainties in cloud feedbacks. For a potential PBL mission, maturation of our proof-of-concept along the lines described in Section 3 could lead to breakthroughs in understanding mission value by rigorously connecting architecture proposals to reductions in prediction uncertainties with process-oriented causal attribution. By extension, if a planned PBL mission were descoped, the impact of that loss could be assessed. More generally, an ESM-based climate OSSE approach will be relevant whenever a planned observing system has unquantified potential to meaningfully narrow the uncertainty in numerical prediction of societally impactful climate quantities—and the impacts of any future losses or enhancements to planned measurements could be similarly quantified.

From an economic standpoint, while the climate OSSE proof-of-concept offers a means of objectively evaluating and comparing differing observing systems insofar as their capability to reduce uncertainty in climate model projections, it falls short of quantifying a return on investment (ROI). Long-term economic investments and their value to society are particularly sensitive to uncertainty, especially when current decisions that affect future economic damages may be irreversible (Mäler and Fisher 2005). In principle, any climate impact (e.g., sea level rise or extremes in heat and precipitation) can be potentially associated with an economic impact if that economic impact is a calculable function of the frequency or intensity of the climate impact (e.g., heat waves and late frost in Fig. 1). If a monetary value for reduced uncertainty in any particular quantity has been quantified, e.g., trillions of dollars associated with halving uncertainty in ECS (Hope 2015), then a mission ROI can by extension be estimated as a multiple of the OSSE-quantified reduction in that uncertainty.

For major satellite missions targeting improvement in climate model physics, the total ROI from a climate OSSE study itself is almost certainly large; however, policymakers should still be motivated by the degree to which any given improved Earth observations are the best use of resources relative to other possibilities. What economists call the ‘marginal value’ of an investment (i.e. the expected payoff from an incremental increase) is often the optimal way to choose where to commit future resources. Marginal values can generally be deduced from prices;

however, something that has no price (like the environment) makes it necessary to infer societal values from other behaviors (Bockstael and Freeman 2005). Observations that reduce uncertainty in future climate change likely have a high marginal value, that, if properly communicated, could lead to substantial changes in behavior (Crochemore et al. 2024). Moreover, an initial burst of learning can be extremely valuable because it may help rule out the most extreme scenarios (Kelly and Tan 2015), and, as the above analysis suggests, the climate focused OSSE has potential to improve our ability to do that.

We advocate to extend the climate OSSE framework in a manner that would enable economic valuation of future Earth observing investments, but do not minimize the challenge. Economists also have a culture of model intercomparison, and research into the societal impacts of climate change necessarily inherits the uncertainties of the parameters in models of climate change. Layered on top of that is the additional uncertainty from the chaotic and evolving societal responses, with unpredictability that, like climate change itself, can be on the time scale of decades or even centuries. Still the two systems – economic and climatic – will be increasingly intertwined and should be considered together (Bolin 2003).

As discussed above, climate OSSEs should be performed systematically by multiple climate models in order to span structural uncertainties and draw the most robust conclusions for decision support. The US climate modeling enterprise is well-positioned to implement such a multi-model framework by leveraging the diverse and complementary strengths of its six major federally-funded global modeling efforts (Mariotti et al. 2024). All six centers have participated for the last decade in the Interagency Group on Integrative Modeling (IGIM), which organizes annual workshops and summit meetings dedicated to knowledge-sharing and synergy-building; an IGIM-coordinated analysis of model tuning protocols (Schmidt et al. 2017) inspired the GISS-E3 CPE framework, providing a foundation for the climate OSSE approach proposed here.

Of the six centers, NASA's GISS, NOAA's Geophysical Fluid Dynamics Laboratory (GFDL), NSF's National Center for Atmospheric Research (NCAR) and DOE's multi-lab Energy Exascale Earth System Model (E3SM) efforts share a focus on climate modeling and research activities, but are diverse in the specific expertise they can contribute to an OSSE framework. For instance, GISS uniquely focuses on integrating NASA Earth observations into its climate model development and calibration process (Elsaesser et al. 2025) and specializes in the development and application of observation simulators. Thus, GISS is able to provide the supporting infrastructure and technical background on which the proposed climate OSSE framework is built. GFDL, in support of NOAA's broad prediction and conservation mandates, develops models that span the weather, climate, ocean and marine domains and thus has unique capability to quantify the value of an observing platform to a wide variety of natural, economic and societal systems. NCAR, benefiting from its position as the main US developer of open-source community models, can contribute leading expertise in a variety of state-of-the-art tools relevant to this proposed framework, such as advancements in the implementation of perturbed parameter ensembles and the incorporation of ML techniques. DOE's E3SM efforts include a focus on human-Earth system feedbacks and projecting climate impacts on US energy supply and demand. Therefore, the implementation of the proposed framework by DOE would greatly expand the relevance of climate OSSEs to actionable decision making, and can also diversify the types of observing platforms that such OSSEs can be used to study.

The US climate modeling enterprise also includes two centers with a decades-long focus on implementing already-established OSSE frameworks: NASA's Global Modeling and Assimilation Office (GMAO) and NOAA's Environmental Modeling Center (EMC). While their technical expertise in OSSE development may overlap, they use OSSEs for different purposes and provide decision support to US agencies with considerable programmatic differences. For instance, GMAO supports development of research-oriented observing platforms, while EMC largely supports operational measurements and their assimilation into NWP. EMC's OSSE efforts are also largely congressionally mandated, falling under the Weather Research and Forecasting Innovation Act of 2017. These differing experiences will be valuable for determining how climate OSSEs can best serve decision making at the agency level.

## 5. Summary and conclusions

We outline a new ESM-based climate OSSE approach for supporting satellite mission design in cases where the data are intended to better constrain model physics. The approach relies on a new ML-enabled method for objective calibration of the NASA GISS-E3 climate model to a diversity of existing satellite data sets, which yields multiple plausible combinations of uncertain parameters, referred to as a calibrated physics ensemble (CPE). Simulations with all CPE versions of the GISS-E3 climate model in turn yield an envelope of climate predictions, thus quantifying the effects of model physics uncertainty on projections. Since the CPE relies objectively on the existing program of record (POR) satellite data and its uncertainty characteristics, it provides a vehicle to quantify how a hypothetical improvement in the POR would change the GISS-E3 CPE and its predictions. We illustrate this by demonstrating a rederivation of the GISS-E3 CPE for a hypothetical reduction of uncertainties in future near-surface qv925 and T925 satellite data that could be provided by a NASA PBL mission. Results of this simple demonstration suggest that improved qv925 measurements could reduce uncertainty in GISS-E3 projections in a manner not matched by improved T925 measurements, qualitatively consistent with a single-column model study (Suselj et al. 2020).

Development of this proof-of-concept into a mature capability on a par with NOAA QOSAP OSSE tools would benefit from a number of extensions. Using more than one climate model would allow consideration of structural uncertainty, which was not examined here. The addition of Bayesian network metrics could support process-oriented causal attribution of the improvements to specific measurement aspects, which was not attempted here. Additional attention to climate projection metrics and uncertainty quantification would support both mission design and economic valuation (ROI). In the US, modeling centers are well-positioned to support climate OSSE studies. We advocate for appending basic economic valuations of future Earth observing investments within the climate OSSE framework.

Uncertainties in ESM climate change projections limit our ability to quantify the extent and impact of climate change on multidecadal time scales. Reducing these uncertainties with current or future observations is an important priority for agencies with Earth observing programs world-wide. A climate OSSE framework such as described here could contribute substantially to this goal, both through enabling more effective utilization of the current observing systems and as a means of identifying critical observations for future missions. In the US, planning for future Earth observing satellite missions is strongly informed by the Decadal Survey process, which would benefit from the inclusion of climate OSSE studies. The climate OSSE approach described here is also suitable for any stage of mission proposal and development, and could readily be implemented in potentially limited forms such as one or two-model studies (e.g., Chepfer et al. 2018) to guide mission priorities.

*Acknowledgments.*

A portion of this research was conducted at the Jet Propulsion Laboratory, California Institute of Technology, under a contract with the National Aeronautics and Space Administration (NASA) 80NM0018D0004. GSE and MvLW acknowledge additional support from the NSF STC Learning the Earth with Artificial Intelligence and Physics (LEAP) (NSF Award Number 2019625). MDZ's work was supported by the U.S. Department of Energy (DOE) Regional and Global Model Analysis program area and was performed under the auspices of the U.S. DOE by Lawrence Livermore National Laboratory under Contract DEAC52-07NA27344.

*Data Availability Statement.*

The data needed for reproduction of Figs. 2 and 3 are available at https://doi.org/10.5281/zenodo.15116918 (Elsaesser, 2025).